\documentclass[twocolumn,showpacs,amsmath,amssymb,superscriptaddress,floatfix]{revtex4}

\usepackage{graphicx}
\usepackage{dcolumn}
\usepackage{bm}
\usepackage{multirow}
\usepackage{indentfirst}
\usepackage{rotating}

\begin{document}
\title{The energy dependence of the $\boldsymbol{pp \to K^+ n \Sigma^+}$ reaction close to threshold}

\author{Yu.~Valdau}%
\email{y.valdau@fz-juelich.de}%
\affiliation{High Energy Physics Department, Petersburg Nuclear
Physics Institute, RU-188350 Gatchina, Russia}%
\affiliation{Institut f\"ur Kernphysik and J\"ulich Centre for Hadron Physics,
 Forschungszentrum J\"ulich, D-52425 J\"ulich, Germany}
\author{S.~Barsov}
\affiliation{High Energy Physics Department, Petersburg Nuclear
Physics Institute, RU-188350 Gatchina, Russia}%
\author{M.~B\"uscher}
\affiliation{Institut f\"ur Kernphysik and J\"ulich Centre for Hadron
Physics, Forschungszentrum J\"ulich, D-52425 J\"ulich, Germany}
\author{D.~Chiladze}
\affiliation{Institut f\"ur Kernphysik and J\"ulich Centre for Hadron
Physics, Forschungszentrum J\"ulich, D-52425 J\"ulich, Germany}
\affiliation{High Energy Physics Institute, Tbilisi State University,
0186 Tbilisi, Georgia}
\author{S.~Dymov}
\affiliation{Physikalisches Institut II, Universit{\"a}t
Erlangen-N{\"u}rnberg, D-91058 Erlangen, Germany }
\affiliation{Laboratory of Nuclear Problems, JINR, RU-141980 Dubna,
Russia}
\author{A.~Dzyuba}
\affiliation{High Energy Physics Department, Petersburg Nuclear
Physics Institute, RU-188350 Gatchina, Russia}%
\author{M.~Hartmann}
\affiliation{Institut f\"ur Kernphysik and J\"ulich Centre for Hadron Physics,
 Forschungszentrum J\"ulich, D-52425 J\"ulich, Germany}
\author{A.~Kacharava}
\affiliation{Institut f\"ur Kernphysik and J\"ulich Centre for Hadron Physics, 
Forschungszentrum J\"ulich, D-52425 J\"ulich, Germany}
\author{I.~Keshelashvili}
\affiliation{Physics Dept., University of Basel, Klingelbergstrasse
82, CH-4056 Basel, Switzerland}
\affiliation{High Energy Physics Institute, Tbilisi State University, 0186 Tbilisi, Georgia}
\author{A.~Khoukaz}
\affiliation{Institut f\"ur Kernphysik, Universit\"at
M\"unster, D-48149 M\"unster, Germany}%
\author{V.~Koptev}%
\affiliation{High Energy Physics Department, Petersburg Nuclear
Physics Institute, RU-188350 Gatchina, Russia}
\author{P.~Kulessa}
\affiliation{H.\,Niewodniczanski Institute of Nuclear Physics PAN,
PL-31342 Cracow, Poland}
\author{S.~Merzliakov}
\affiliation{Laboratory of Nuclear Problems, JINR, RU-141980 Dubna,
Russia} \affiliation{Institut f\"ur Kernphysik and J\"ulich Centre
for Hadron Physics, Forschungszentrum J\"ulich, D-52425 J\"ulich,
Germany}
\author{M.~Mielke}
\affiliation{Institut f\"ur Kernphysik, Universit\"at
M\"unster, D-48149 M\"unster, Germany}%
\author{S.~Mikirtychiants}
\affiliation{High Energy Physics Department, Petersburg Nuclear
Physics Institute, RU-188350 Gatchina, Russia}
\affiliation{Institut f\"ur Kernphysik and J\"ulich Centre for Hadron Physics, 
Forschungszentrum J\"ulich, D-52425 J\"ulich, Germany}
\author{M.~Nekipelov}
\affiliation{Institut f\"ur Kernphysik and J\"ulich Centre for Hadron
Physics, Forschungszentrum J\"ulich, D-52425 J\"ulich, Germany}
\author{H.~Ohm}
\affiliation{Institut f\"ur Kernphysik and J\"ulich Centre for Hadron
Physics, Forschungszentrum J\"ulich, D-52425 J\"ulich, Germany}
\author{M.~Papenbrock}
\affiliation{Institut f\"ur Kernphysik, Universit\"at
M\"unster, D-48149 M\"unster, Germany}%
\author{F.~Rathmann}
\affiliation{Institut f\"ur Kernphysik and J\"ulich Centre for Hadron
Physics, Forschungszentrum J\"ulich, D-52425 J\"ulich, Germany}
\author{V.~Serdyuk}
\affiliation{Laboratory of Nuclear Problems, JINR, RU-141980 Dubna,
Russia}
\affiliation{Institut f\"ur Kernphysik and J\"ulich Centre for Hadron Physics, 
Forschungszentrum J\"ulich, D-52425 J\"ulich, Germany}
\author{H.~Str\"oher}
\affiliation{Institut f\"ur Kernphysik and J\"ulich Centre for Hadron
Physics, Forschungszentrum J\"ulich, D-52425 J\"ulich, Germany}
\author{S.~Trusov}
\affiliation{Institut f\"ur Kern- und Hadronenphysik,
Forschungszentrum Rossendorf, D-01314 Dresden, Germany}
\affiliation{Skobeltsyn Institute of Nuclear Physics, Lomonosov Moscow 
State University, RU-119991 Moscow, Russia}
\author{C.~Wilkin}
\affiliation{Physics and Astronomy Department, UCL, London WC1E 6BT,
United Kingdom}

\date{\today}%

\begin{abstract}
The production of the $\Sigma^+$ hyperon through the $pp\to K^+ n
\Sigma^+$ reaction has been investigated at four energies close to
threshold, 1.826, 1.920, 1.958, and 2.020~GeV. At low energies, 
correlated $K^+\pi^+$ pairs can only originate from $\Sigma^+$
production so that their measurement allows the total cross section
for the reaction to be determined. The results obtained are completely
consistent with the values extracted from the study of the
$K^+$-proton correlation spectra obtained in the same experiment. These
spectra, as well as the inclusive $K^+$ momentum distributions, also provide
conservative upper limits on the $\Sigma^+$ production rates. The measurements show
a $\Sigma^+$ production cross section that varies roughly like phase
space and, in particular, none of the three experimental approaches
used supports the anomalously high near-threshold $pp\to K^+ n
\Sigma^+$ total cross section previously reported [T.~Ro\.zek
\textit{et al.}, Phys.\ Lett.\ B \textbf{643}, 251 (2006)].
\end{abstract}


\maketitle

%
%
\section{Introduction}
\label{Introduction}%

The energy dependence of the total cross section for associated
strangeness production in the exclusive $pp\to K^+p\Lambda$ reaction
near threshold has been investigated in a series of experiments
undertaken at the COSY accelerator of the Forschungszentrum J\"ulich
using the COSY-11 and COSY-TOF
detectors~\cite{BAL96,BIL98,SEW99,EYR03,KOW04,SAM06}. The resulting
behavior can be described in terms of a three-body phase space that
is influenced by a strong final state interaction (FSI) between the
$\Lambda$ hyperon and the emerging proton.

The situation for $\Sigma^0$ production in the analogous $pp\to
K^+p\Sigma^0$ reaction is rather different since the existing data
show a total cross section that varies as expected from three-body
phase space, with no evidence for a strong $\Sigma^0p$
FSI~\cite{SEW99,KOW04,VAL07}. A second distinction between the
production of the two hyperons in proton-proton collisions is that,
at the same value of the excess energy $\varepsilon$ close to
threshold, the total cross section for $pp\to K^+p\Sigma^0$ is at
least one order of magnitude less than that of $pp\to K^+p\Lambda$.
This probably reflects differences in the underlying reaction
mechanisms as well as between the two FSI involved.

Since the $\Sigma$ has isospin $I=1$, extra light may be cast on
 both the FSI and reaction mechanism questions through the study of
$\Sigma^+$ production in the $pp\to K^+n\Sigma^+$ reaction. The 
first measurements on this channel were completed by the COSY-11
collaboration at proton beam energies of $T_p=1.826$~GeV and
1.958~GeV, \emph{i.e.}, at $\varepsilon = 13$~MeV and 60~MeV, by
detecting the neutron in coincidence with the produced
$K^+$~\cite{ROZ06}. The values obtained for the total cross section
were strikingly high compared to those for $\Sigma^0$ production, or
indeed with those for $\Lambda$ production, with ratios of
$R(\Sigma^+/\Sigma^0)= \sigma(pp\to K^+n\Sigma^+)/ \sigma(pp\to
K^+p\Sigma^0) =230{\pm}70$ at $\varepsilon = 13$~MeV and $90{\pm}40$
at 60~MeV~\cite{ROZ06}. Such values would imply a very anomalous
energy dependence, since bubble chamber data yield a ratio of
$R(\Sigma^+/\Sigma^0)=2.3\pm 0.9$, though at the much higher excess
energy of $\varepsilon = 350$~MeV~\cite{LOU61}.

In an effort to understand the situation, a first measurement of the
$pp\to K^+n\Sigma^+$ reaction was carried out at the COSY-ANKE
facility that did not rely on the detection of the final
neutron~\cite{VAL07}. This experiment at $\varepsilon=129$~MeV
($T_p=2.16$~GeV) involved a three pronged approach. Although the
statistics were low, apart from a small $K^+n\Lambda\pi^+$
background, the detected $K^+\pi^+$ coincidences at this energy could
 only have come from $K^+n\Sigma^+$, where the hyperon decays
through $\Sigma^+\to n\pi^+$ ($\textrm{BR}=48.3\%$). This approach
indicated that $R(\Sigma^+/\Sigma^0)$ was of the order of unity at
this energy.

Information on the $K^+n\Sigma^+$ channel could also be obtained
through the measurement of $K^+$-proton correlations, where the
$\Sigma^+$ shows up as a contribution in the missing-mass
distribution via its decay $\Sigma^+ \to p\pi^0$
($\textrm{BR}=51.6\%$). Although the overall spectra are sensitive to 
the model used to describe the dominant $K^+p\Lambda$ channel, at
high $K^+p$ missing masses there can only be contributions from
$\Sigma$ production. Both this study and that of the inclusive $K^+$ 
production showed consistency with the $K^+\pi^+$
determination and clearly excluded a large value of $R(\Sigma^+/\Sigma^0)$.

These first ANKE results were obtained at a significantly higher
excess energy than those of COSY-11~\cite{ROZ06} and so did not rule
out the possibility of a strong threshold anomaly. The aim of the
present work is to extend the earlier ANKE methodology by making 
four measurements that cover the near-threshold region as well as one
below the $\Sigma$ threshold to obtain data where the $K^+p\Lambda$
channel is the only source of associated strangeness
production~\cite{VAL09}. The beam energy range covered was from 1.775 to 2.020~GeV.

We start by summarizing in Sec.~\ref{data_and_models} the existing
near-threshold data on hyperon production in proton-proton collisions 
and their interpretation in terms of phenomenological models. The
ANKE experimental facility used for the study of the $pp\to
K^+n\Sigma^+$ reaction and the techniques involved there are
described in Sec.~\ref{Experiment}. Section~\ref{results} presents first
the results of the measurements of inclusive $K^+$ production.
The comparison of these data taken just above with those from just below the
$\Sigma$ thresholds allows us to set an upper limit on the $\Sigma^+$ 
production cross section at $\varepsilon=13$~MeV that is almost two
orders of magnitude smaller than the COSY-11 result~\cite{ROZ06}. 
The $K^+p$ missing-mass spectra lead to values of the total cross sections for
both $\Lambda$ and $\Sigma^0$ production that are in
agreement with the world data. The high ends of these spectra can only be populated 
by events arising from $\Sigma$ production and these allow first estimates to 
be made of the $\Sigma^+$ cross section as well as conservative upper limits.
Although the statistics here are reasonable, it is hard to quantify the 
systematic uncertainties and, in this respect,
the most reliable determination of the $\Sigma^+$ production cross
section is through the study of $K^+\pi^+$ coincidences which, at the
 energies of this experiment, can only arise from $\Sigma^+$ decay.
The values obtained from this study are completely consistent with those found
from the $K^+p$ coincidence data and fall below the upper bounds set 
by the inclusive data.

The results for the total cross sections are discussed in
Sec.~\ref{Discussion}, where they are compared to other 
measurements of $\Lambda$ and $\Sigma^0$ production as well 
as those of $\Sigma^+$. The clean determinations of the 
$pp\to K^+ n\Sigma^+$ cross section from the $K^+\pi^+$ 
and $K^+p$ measurements give values that are more than 
two orders of magnitude lower than those reported in the 
literature through the detection of the neutron~\cite{ROZ06}. 
Our conclusions and outlook for the future are presented in Sec.~\ref{Conclusions}.
%
%
\section{Hyperon production in proton-proton collisions}
\label{data_and_models}%

In the low energy regime (around $T_p\approx 2$~GeV) there are three
associated-strangeness production channels allowed in proton-proton
collisions, viz.\ $K^+p\Lambda$, $K^+p\Sigma^0$, and $K^+n\Sigma^+$.
Measurements of inclusive $K^+$ spectra lead to distributions where
there are contributions from all three final states. There is no
possibility of distinguishing between the different reaction 
channels without modeling the distributions from all of
them~\cite{SIE94,SIB07}, and this can lead to severe ambiguities. 
The basic problem here is that the cross section for $\Lambda$ 
production well away from threshold is much larger than those for the
$\Sigma$ close to threshold and there is no reason to believe that
the associated distributions should vary like phase space.

The conditions improve when two particles are detected in the final
state. Signals from the $pp\to K^+ p \Lambda$ and $pp \to
K^+p\Sigma^0$ reactions are usually identified through the
corresponding peaks in the $K^+p$ missing-mass spectra~\cite{SEW99}.
The physical backgrounds in such measurements are mostly connected
with the decays of $\Lambda$ and $\Sigma^0$ hyperons, which lead 
to a second proton in the final state (see Table~\ref{tab:tRCh}). The
remaining part of $K^+p$ missing-mass spectra can be associated with
the $pp\to K^+n\Sigma^+$ reaction channel, where the $\Sigma^+\to p\pi^0$ 
decay gives rise to the proton that is detected. 
It is important to note that the missing-mass range that is obtained 
using a proton from $\Lambda$ decay does not extend as far as that
linked to $\Sigma$ production so that the study of the maximum missing-mass
region allows one to derive values for the weighted sum of $\Sigma^0$
and $\Sigma^+$ production. Nevertheless, the contributions are rather small 
and systematic effects can be significant near this kinematic limit.

However, below the threshold for $pp\to K^+n\Lambda\pi^0$ production
($T_p=1.975$~GeV), the only source of the $K^+\pi^+$ correlations is
$\Sigma^+$ production and this provides a much more reliable way to
identify the channel than the alternative of detecting the
neutron~\cite{ROZ06}. Although the measurement of $K^+\pi^+$
coincidences provides a very clean signal for the identification of
the $pp\to K^+n\Sigma^+$ reaction, the statistics achievable at ANKE
are quite low~\cite{VAL07,VAL09}. 

\begin{table}[ht]
\begin{center}
\begin{tabular}{|c|c|c|c|c|}
\hline
 \multicolumn{2}{|c|}{Final state}  & $K^+p\Lambda$ & $K^+p\Sigma^0$ & $K^+n\Sigma^+$\\
\hline
\multicolumn{2}{|c|}{$T_{\text{thr}}$ [GeV]}     & 1.582 & 1.794 & 1.789\\
\multicolumn{2}{|c|}{BR($K^+p$) [\%]}     & 63.9  & 63.9  & 51.6\\
\multicolumn{2}{|c|}{BR($K^+\pi^+$) [\%]} & --    & --    & 48.3\\
\hline
\end{tabular}
\end{center}
\caption{Characteristics of the three hyperon production channels
in proton-proton collisions studied in this experiment. In addition
to the threshold beam energies, also shown are the fractions of
the cross sections (BR) associated with the $K^+\pi^+$ and indirect
 $K^+p$ coincidences.
 \label{tab:tRCh}
}
\end{table}

It is evident from the inclusive $K^+$ production data that there is
a strong $\Lambda p$ final state interaction at low invariant
masses~\cite{SIE94,HIRES}. This is reflected in the energy
dependence of the $pp\to K^+p\Lambda$ total cross section analyzed 
in Refs.~\cite{SIB06,SIB07}, where the near-threshold experimental data
are enhanced compared to phase space. Significant influence of the
$\Lambda p$ final state interaction has also been found in the
analysis of $pp\to K^+p\Lambda$ Dalitz plots~\cite{SAM06, EYR03}. 
The measured distributions have been described using a $\Lambda$
production model which, in addition to the strong FSI, includes the
$N^*(1650)$-isobar in the production mechanism. It is believed that
 this is the dominant resonance in the $\approx2$~GeV region~\cite{EYR03}.

The ANKE acceptance for the $pp\to K^+ p \Lambda$ reaction changes
significantly over the energy range where the data were collected. 
In addition, different models have been used in this analysis for the
estimation of the $\Lambda$ total cross section at different
energies. A three-body phase space, modified by the
$\Lambda p$ FSI, has been used for $T_p<1.9$~GeV. Above 1.9~GeV, a
model for the $pp\to K^+ p \Lambda$ reaction, similar to the one 
used for the $2.16$~GeV data~\cite{VAL07}, has been developed. This
assumes the dominance of the $N^*(1650)$-resonance in the
production mechanism and a strong $\Lambda p$ final state 
interaction. In addition to these effects, there is some anisotropy 
in the angular spectra of the kaon and direct proton in the 
$K^+p\Lambda$ final state~\cite{MET98, HES00, FRI02,SCH03}. 
Unfortunately, unlike the case for the $2.16$~GeV data~\cite{SCH03}, 
there are no measurements of the angular distributions directly 
at the energies of interest. The relevant parameters have 
therefore been determined from linear interpolation of the results
 of Refs.~\cite{MET98, HES00, FRI02, SCH03} at nearby energies. 
More details of the model and the parameters used in the Monte 
Carlo simulations can be found in Ref.~\cite{VAL09}.

Extensive measurements of the total cross section for $\Sigma^0$ 
production close to threshold were carried out by the COSY-11
collaboration~\cite{SEW99,KOW04}. Some studies of differential
observables were undertaken at the COSY-TOF facility~\cite{BIL98}, 
though most of the results have only been reported in
theses~\cite{FRI02,SCH03}. There is little evidence from these data
that in the vicinity of threshold there are significant deviations
from phase space. We therefore use a phase-space description when
evaluating the acceptance for $\Sigma^0$ production at ANKE. Since
there is no experimental information at all on differential
quantities for the $pp\to K^+n\Sigma^+$ reaction channel at low
energies, a similar three-body phase-space description has been
employed for $\Sigma^+$ production.

The $pp\to K^0p\Sigma^+$ reaction was studied at
$\varepsilon = 126$~\cite{ABD04} and 161~MeV~\cite{ABD07} during 
the COSY-TOF pentaquark searches. The value of the total cross 
section deduced from these data at the lower energy is quite 
close to that found for the $K^+p\Sigma^0$ final state~\cite{SCH03} 
and this already allows simple bounds to be placed on the cross 
section for the $pp\to K^+n\Sigma^+$ reaction.

There are only two independent isospin amplitudes for $\Sigma$
production in proton-proton collisions and these can be taken to
correspond to the $I=1/2$ and $I=3/2$ combinations of the final
$\Sigma N$ pair. There is therefore a linear relation between the
amplitudes for the production of the three possible final states:
\begin{align}
\nonumber%
& f(pp \to K^+ n \Sigma^+)+ f(pp \to K^0 p \Sigma^+)\\
& + \sqrt{2}\,f(pp \to K^+ p \Sigma^0) = 0\,.
\label{amps}%
\end{align}
This leads to a triangle inequality between the total cross
sections~\cite{LOU61}, which can provide model-independent limits on
the ratio of $\Sigma^+$ to $\Sigma^0$ production:
\begin{align}
\nonumber &\left[\sqrt{\sigma(pp \to K^0 p
\Sigma^+)}-\sqrt{2\sigma(pp
\to K^+ p \Sigma^0)}\,\right]^2\\
\nonumber &\hspace{2cm}\leq \sigma(pp \to K^+ n \Sigma^+)\\
&\leq \left[\sqrt{\sigma(pp \to K^0 p \Sigma^+)}+\sqrt{2\sigma(pp \to K^+ p \Sigma^0)}\,\right]^2.
\label{triangle}%
\end{align}

The COSY-TOF result at $\varepsilon = 126$~MeV~\cite{ABD04} 
suggests that at the energy of the earlier ANKE experiment at 2.16~GeV
the ratio $R(\Sigma^+/\Sigma^0)$ of the total cross sections for
$\Sigma^+$ to $\Sigma^0$ production should lie in the range $1/6
<R(\Sigma^+/\Sigma^0)<6$. Though weak, this bound is an order of
magnitude smaller than that found by the COSY-11 collaboration,
though at lower energies~\cite{ROZ06}. A direct measurement of the
$pp\to K^+n\Sigma^+$ cross section at this energy using the COSY-TOF
spectrometer~\cite{KAR05} satisfies well this inequality, whereas an
earlier one at $\varepsilon =94$~MeV seems to be much too
large~\cite{SCH03a}.

%
%

\section{Experiment}
\label{Experiment}%
The experiment was carried out at the ANKE facility~\cite{BAR01}
with an unpolarized proton beam provided by the Cooler Synchrotron 
and storage ring COSY~\cite{MAI97}. Of the five beam energies used,
$T_p=1.775$, 1.826, 1.920, 1.958 and 2.020~GeV, the first, just
below the threshold for the production of the $\Sigma$ hyperon,
was used for background studies. The values of the excess energies 
$\varepsilon$ for the three final states are given in Table~\ref{tab:ExEn}.

\begin{table}[ht]
\begin{center}
\begin{tabular}{|c|c|c|c|c|}
\hline
\multicolumn{2}{|c|}{Final state}  & $K^+p\Lambda$ & $K^+p\Sigma^0$ & $K^+n\Sigma^+$\\
\hline
$p$     & $T_p$ & \multicolumn{3}{c|}{$\varepsilon$}\\
GeV/$c$ & GeV   & \multicolumn{3}{c|}{MeV}\\
\hline
2.546 & 1.775 & 70                  & --                   &--\\
2.600 & 1.826 & 88                  & 11                   &13\\
2.700 & 1.920 & 122                 & 45                   &47\\
2.740 & 1.958 & 135                 & 58                   &60\\
2.806 & 2.020 & 157                 & 80                   &82\\
\hline
\end{tabular}
\end{center}
\caption{ Values of the excess energies $\varepsilon$ for the three
reaction channels at the five beam momenta $p$ and energies $T_p$ used in this experiment. \label{tab:ExEn} }
\end{table}

%
%
\begin{figure*}[hbt]
\begin{center}
\includegraphics[scale=0.75]{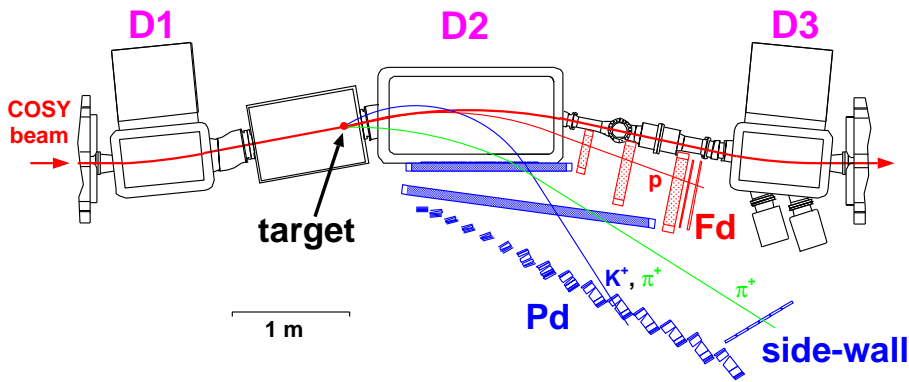}
\end{center}
\caption{(Color online) Sketch of the relevant parts of the ANKE
detector system showing the positions of the two bending magnets
$D1$ and $D3$ and the target before the analyzing magnet $D2$.
Only information from the forward (Fd) and positive side (Pd)
detectors, including the stop counters in the side wall, was used in this experiment.
\label{fig:fAnke}}
\end{figure*}
%
%

The ANKE magnetic spectrometer, located at an internal target
position of COSY, is composed of three dipole magnets. $D1$ and $D3$
bend the direct proton beam from the undisturbed COSY orbit to the
ANKE target and return it back, respectively. The analyzing magnet
$D2$, placed between $D1$ and $D3$, was operated in this experiment
with a field strength of $B=1.57$~T. This magnet deflects positively
charged particles with laboratory polar angles up to $20^{\circ}$
towards the positive side (Pd) and forward (Fd) detector systems. 
The high density of the unpolarized H$_2$ cluster-jet target~\cite{KHO99} 
gave a typical target thickness of $\sim10^{14}~$atoms/cm$^{2}$ 
and this resulted in an average luminosity of the order of
$\sim2~$nb$^{-1}$s$^{-1}$. In the present analysis only the positive
side (for the $\pi^+$, $K^+$ and $p$ identification) and forward
detector systems (for proton identification) have been used. 
A sketch of the ANKE detector system is presented in Fig.~\ref{fig:fAnke}.

The positive side detector system consists of 23 start (Sa) and 21
stop (So) counters for time-of-flight (TOF) measurements and two
multiwire proportional chambers (MWPC) for track reconstruction and
background suppression. Depending on the type and momentum of the
particle, the tracking efficiencies of the Pd MWPCs are typically
$\sim90$\%, the values being determined using experimental data. The
momentum resolution in the positive side detector system is about 2\% (FWHM).

The first fifteen stop counters are part of the range telescope
system designed for the identification of low momentum $K^+$
($p_K^{\text{max}}\approx 620$~MeV/$c$ for $B=1.57$~T)~\cite{BUS02}.
The remaining six stop counters, joined together to form the
side-wall detector ($p^{\text{max}}\approx 920$~MeV/$c$ for
$B=1.57$~T), were used for $\pi^+$ and $p$ identification on the
basis of time-of-flight criteria.

Each of the range telescopes is made up of a stop counter, an energy
loss counter ($\Delta E$), a so-called delayed veto counter, and two
passive degraders made of copper. The thickness of the first degrader 
is chosen such that a $K^+$ deposits all its energy in the $\Delta E$ 
counter and stops either at the edge of it or in the second degrader. 
The products from the $K^+$ decay are registered in the delayed veto 
counter with the characteristic decay time of $12.4$~ns. All the range 
telescopes, placed in the focal plane of the $D2$ magnet, are optimized 
for the measurement of a given $K^+$ momentum. The delayed veto criteria 
leads to a suppression of better than $10^{-5}$ in the non-kaon 
background for both inclusive and coincidence measurements. The details 
of the $K^+$ identification using the delayed veto technique can be 
found in Ref.~\cite{BUS02}.

Due to the unchanged $D2$ magnetic field, the geometry of the
positive side detector system was fixed with respect to $D2$ 
so that the acceptance of each
telescope remained the same at all the beam energies
where experimental data were collected. As a consequence, the same
set of detector efficiencies and acceptances could be used in the
analysis of the inclusive data collected at different energies.

The ANKE forward detector system used for the $K^+p$ correlation
measurements and the luminosity determination consists of two
multiwire proportional chambers, one drift chamber, and a hodoscope
of scintillator counters~\cite{DYM04}. The tracking efficiency of 
the Fd chambers, as determined from the experimental data, is 
homogeneous over the Fd acceptance and is better than 95\% for 
protons. The ensemble of
chambers used in the experiment allows the reconstruction of the
proton momentum with a precision of $\sim2$\% (FWHM). The Fd
hodoscope comprises two layers of plastic scintillator counters
shifted with respect to each other by half the width of a counter.
The two-layer structure allows the time from individual counters to
be calibrated with respect to each other by using tracks of 
particles which cross counters that overlap in different layers. The time
signals from individual forward counters are therefore aligned such
that in the analysis the full hodoscope can be treated as a single
counter~\cite{DYM04}.

In order to normalize the experimental data, and hence extract
absolute cross sections for $\Sigma^+$ production, the proton-proton
elastic scattering rate was measured using the Fd detector. Due to
the high count rate from this reaction channel, a dedicated trigger,
prescaled by a factor of a thousand, was used. Having determined the
momentum of a proton candidate in the Fd, the $pp\to pp$ reaction
was identified via the missing-mass technique. The elastic
scattering peak, with a width of $50$~MeV/$c^2$ (FWHM), is well
separated from the inelastic background arising from pion production. 
The numbers of protons scattered between laboratory angles of
$6^{\circ}$ and $9^{\circ}$ were determined from a fit to the
data by a Gaussian with polynomial background. Predictions for the
$pp$ elastic differential cross section from the SAID
analysis~\cite{ARN00} were then used to evaluate the luminosity. For
all the beam energies in the experiment, the shape of the differential
 cross section as a function of laboratory angle agreed with the 
predictions to better than 3\%. The luminosity was determined 
with an overall uncertainty of 7\%. This takes into account a 5\% 
estimate of the uncertainty from the SAID program, 3\% uncertainty 
in the analysis algorithm, and $3$\% uncertainty in the Fd 
MWPC efficiency determination. The values of the total luminosities 
accumulated are reported in Table~\ref{tab:tKPiInfo}.

The Runge-Kutta method of momentum reconstruction was employed for
the final analysis. This uses all the information from the chambers
and the measured $D2$ magnet field map~\cite{DYM00}. The precise
knowledge of the particle trajectories provided by this method 
allows one to calibrate absolutely the times between different 
start-stop counters in the Pd using experimental data with a 
$\pi^+$ or any other cleanly selected particle. The internal 
delays for time signals do not change during the experiment 
and the required constants could be calculated using time information 
from individual counters and corresponding tracks. After performing 
the time calibration, the invariant mass of a particle with a 
given track could be calculated.

\begin{figure}[htb]
\begin{center}
\includegraphics[scale=0.4]{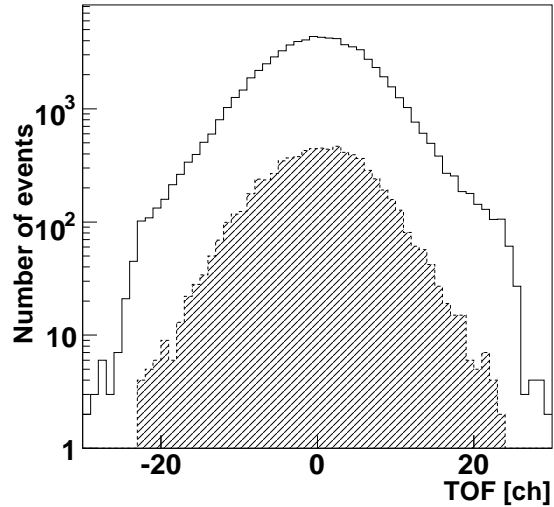}
\end{center}
\caption{Time-of-flight (TOF) spectra, measured at 1.920~GeV with a
delayed veto, presented on a logarithmic scale. The data, measured
in units of 44~ps per channel, have been centered around zero.
The hatched histogram shows the distributions of $K^+$ events for
which there were $K^+p$ coincidences. \label{fig:fTof}}
\end{figure}

The time-of-flight spectra measured at 1.920~GeV for $K^+$ selected using the
delayed veto technique are presented in Fig.~\ref{fig:fTof} summed
over all stop counters.
%
%
%
%

For the extraction of the inclusive double-differential cross
section, the residual background was subtracted from the
time-of-flight spectra measured in individual stop counters. The
number of $K^+$ detected in an individual telescope, $N_{K^+}$, has
been determined using a fit of the TOF spectra that assumes a
Gaussian peak from the kaons and a flat background. The
background contribution to the individual stop counter
distributions is less than 5\%.

The laboratory cross section has been evaluated for every
momentum bin from
\begin{equation}
\frac{d^{\,2}\sigma_{K^+}}{d\Omega\,dp}(T_p)=
\frac{N_{K^{+}}}{\Delta{p}\,\Delta\Omega}
\frac{1}{L_{\text{tot}}\,\epsilon_{K^{+}}}, \label{eq:d2sigmadOmdP}
\end{equation}
where $\Delta\Omega$ is the solid angle integration region and
$L_{\text{tot}}$ the integrated luminosity. A momentum bin of
$\Delta{p}= 24$~MeV/$c$ was imposed in the course of the data
analysis. The efficiency of $K^+$ identification, $\epsilon_{K^{+}}$, is estimated on the basis of
\begin{equation}
\epsilon_{K^+} =
\epsilon^{\text{tel}}\times\epsilon^{\text{scint}}\times
\epsilon^{\text{MWPC}}\times\epsilon^{\text{acc}}\,.
 \label{eq:effK}
\end{equation}

Due mainly to geometric acceptance factors, the delayed-veto
technique leads to a telescope efficiency $\epsilon^{\text{tel}}$ of
about 30\% for low momentum kaons but only 10\% for the highest
momenta~\cite{BUS02}. The value for each telescope has been
determined with a precision of 5\% by using $K^+p$ correlations
recorded with a dedicated trigger. The scintillator counter
efficiency $\epsilon^{\text{scint}}$ is higher than 98\% while that
of the MWPC, $\epsilon^{\text{MWPC}}$, varies between $90-95$\%,
depending on the $K^+$ momentum. These efficiencies have been
determined using experimental data. The acceptance correction factor
$\epsilon^{\text{acc}}$ has been estimated using a
GEANT4~\cite{AGO03} model of the ANKE spectrometer. The overall 
$K^+$ detection efficiency $\epsilon_{K^+}$ in this measurement is
independent of beam energy.

The $K^+p$ missing-mass spectra, which depend on the data measured
with the Fd detector, have been used to extract cross sections for
the $\Lambda$ and $\Sigma^0$ production channels. For the selection
of $K^+p$ pairs detected in the forward and positive detectors, the
time differences between the $K^+$ in the Pd and the
proton in the Fd have been calculated. A $3\sigma$ cut has been
applied to subtract the residual background from accidental
coincidences. The background in the $K^+p$ coincidence spectra
measured with the delayed veto is less than 2\%.

%
%
\begin{figure*}[hbt]
\begin{center}
\includegraphics[scale=0.80]{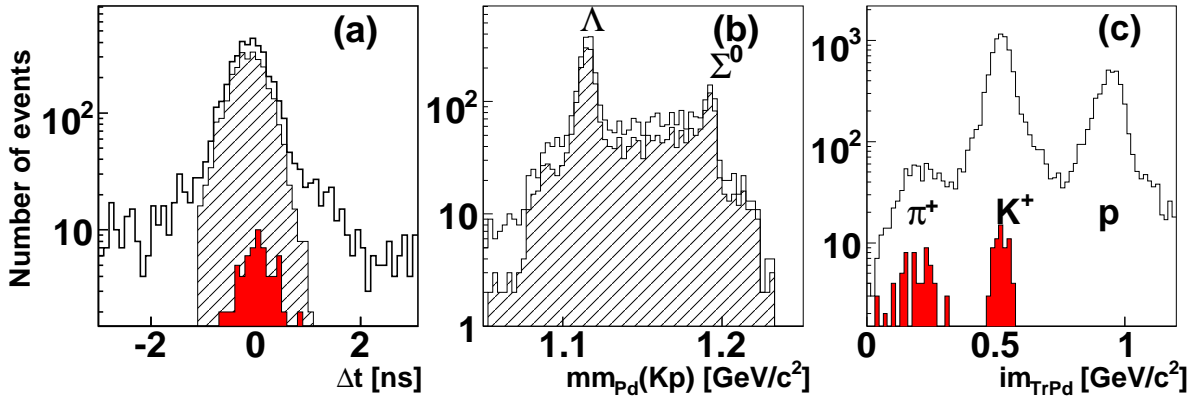}
\end{center}
\caption{(Color online) Identification of two-particle events 
in the positive side detector at 1.920~GeV, where a cut of 
$\theta_{K^+}<12^{\circ}$ has been imposed on the $K^+$ angle. 
(a) The difference $\Delta t$ between measured and calculated 
times of the correlated particles. The selection of double 
coincidences in Pd is shown by the hatched histogram for $K^+p$ 
and the filled histogram (red) for $K^+\pi^+$ pairs. 
(b) The $K^+p$ missing-mass spectra measured in Pd. The hatched 
histogram represents $K^+p$ missing mass selected using the 
cuts shown in panel (a). 
(c) The invariant mass of a detected particle before and after 
(filled histogram) the selection of $K^+\pi^+$ coincidences 
using the criteria in panel (a). \label{fig:fTimeCalib}}
\end{figure*}
%
%

The $K^+$ time-of-flight spectrum measured in coincidence with the
proton detected in the Fd at 1.920~GeV is shown in Fig.~\ref{fig:fTof} 
by the hatched histogram. The overall reduction between the $K^+$ TOF 
spectra measured with and without the proton being detected in the Fd 
is due to the acceptance of this detector. The numbers of $\Lambda$ 
and $\Sigma^0$ events, $N_{\text{ev}}$, in the spectra were estimated 
using fits of the simulations to the $K^+p$ missing-mass spectra. 
The total cross sections were then evaluated from
\begin{equation}
\sigma_{\text{total}}=\frac{N_{\text{ev}}}{L_{\text{total}}\,
\epsilon^{\text{sum}}}\,, \label{eq:totcr}
\end{equation}
where $L_{\text{tot}}$ is the integrated luminosity,
$\epsilon^{\text{sum}}$ the overall efficiency that includes the
detector efficiencies, the in-flight $K^+$ decay correction, and the
total acceptance within a particular reaction model, as
discussed in Sec.~\ref{data_and_models}.

The rate of $K^+ \pi^+$ coincidences is very small. Therefore, the
$K^+p$ correlations detected in the positive detector have been used
to optimize the cut parameters and selection criteria. The $K^+$ and
$p$ tracks were selected using time-of-flight criteria. In
Fig.~\ref{fig:fTimeCalib}a the difference between calculated and
measured time in the So counters is shown, with its center shifted to 
zero. Also presented in panels~b and c are the distributions in
missing mass, assuming that tracks are from $K^+p$ coincidences, and
the invariant mass of the particles whose tracks are detected in the
Pd. The hatched histogram in Fig.~\ref{fig:fTimeCalib}b represents
$K^+p$ coincidences selected using a $3\sigma$ cut on the time
difference shown in panel a. The number of $\Lambda$ and $\Sigma^0$
events, estimated using the clear peaks in the missing-mass
distribution, hardly change. This indicates that the efficiency of
the cuts is better then $\sim97$~\%.

When the same cuts on the time difference are applied to the
$K^+\pi^+$ candidates, the results found are shown in
Figs.~\ref{fig:fTimeCalib}b and c by filled histograms. Since it is
impossible to get correct time information for events where both
particles are detected in the same counter, these are dropped from the
 analysis. This effect is most serious for the $K^+\pi^+$ coincidences
 at lowest energy but even there it does not exceed 25\%. An additional 
correction factor has been introduced to compensate for the loss.

The number of $K^+\pi^+$ correlated pairs has been used to evaluate
the total cross section for $\Sigma^+$ production on the basis of
Eq.~\eqref{eq:totcr}. This includes a correction for the
corresponding branching ratio (48.3\%) reported in
Table~\ref{tab:tRCh}. The acceptance calculation, as well as the
reaction models used, are discussed in Sec.~\ref{data_and_models} 
and given in more detail in Ref.~\cite{VAL09}.

%
%
\section{Experimental results}
\label{results}
%
%
\subsection{Inclusive counting rates}
\label{Inclusive_counts}

\begin{figure}[h!]
\begin{center}
\includegraphics[scale=0.45]{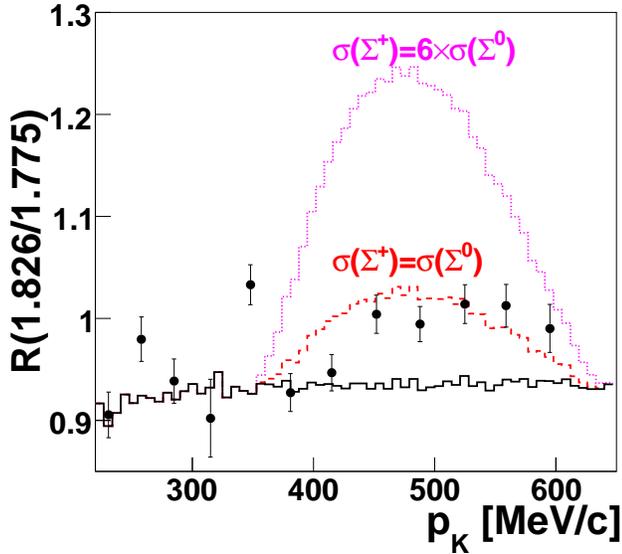}
\end{center}
\caption {(Color online) Ratio of normalized count rates for inclusive 
$K^+$ production at 1.826~GeV to those at 1.775~GeV as a function of 
the momenta in the different telescopes. The data were taken using the 
ANKE range telescope system with $\vartheta_{K^{+}}<12^{\circ}$ and a 
12~MeV/c momentum cut for each telescope. The black solid histogram 
represents the simulation for the $pp\to K^+p\Lambda$ reaction, whereas 
the dashed red one includes also contributions from $\Sigma^+$ and 
$\Sigma^0$ production, assuming $\sigma(\Sigma^0)=\sigma(\Sigma^+)=0.021~\mu$b,
$\sigma_{1.826}(\Lambda)=7.9~\mu$b and $\sigma_{1.775}(\Lambda)=6.1~\mu$b.
The upper magenta dotted histogram represents a simulation where the 
$\Sigma^+$ total cross section is taken to be six times that of $\Sigma^0$,
 which is the limit allowed by the triangle inequality at $\varepsilon = 129$~MeV.
\label{fig:Ratio}}
\end{figure}

A first limit on the cross section for $\Sigma^+$ production can be
obtained by comparing the inclusive $K^+$ counting rates just above
the $\Sigma$ thresholds with those just below. Figure~\ref{fig:Ratio} 
shows the ratio of the data obtained at 1.826~GeV to those at
1.775~GeV as a function of the kaon laboratory momentum. In this way
any questions regarding telescope efficiencies and acceptances are
avoided. At 1.826~GeV the $K^+$ mesons associated with $\Sigma^+$
production have a maximum laboratory angle of
$\vartheta_{K^{+}}^{\text{max}}=12.2^{\circ}$ and a momentum range
from 352 to 643~MeV/c. Essentially the whole phase space for $\Sigma$ 
production is therefore sampled in the data, though the vertical
angular limitation and, to some extent, the horizontal angle cut of
$12^{\circ}$ and the $\pm12$~MeV/c momentum cut for each telescope 
reduces the acceptance to about 23\%. The simulations discussed in
Sec.~\ref{data_and_models} suggest that the contribution to the ratio 
from $\Lambda$ production should be essentially constant over the
whole momentum range, independent of any assumptions made regarding
the kaon angular distributions.

The ratio of differential cross sections can be expressed by Eq.~\eqref{rat}:
\begin{equation}
\label{rat} R\left(\frac{1.826}{1.775}\right)
=\frac{\Lambda_{1.826}}{\Lambda_{1.775}}
+\frac{\Sigma^0_{1.826}+\Sigma^+_{1.826}}{\Lambda_{1.775}}
\end{equation}
The shape expected for the contribution from the combined $pp\to K^+p\Sigma^0$ and $pp\to
K^+n\Sigma^+$ channels is indicated, using total cross sections
$\sigma(\Sigma^0)=\sigma(\Sigma^+)=0.021~\mu$b,
$\sigma_{1.826}(\Lambda)=7.9~\mu$b and $\sigma_{1.775}(\Lambda)=6.1~\mu$b.
Depending upon how the $\Lambda$ level is drawn, this comparison
gives a cross section ratio of $(\sigma(pp\to K^+p\Sigma^0)+\sigma(pp\to K^+n\Sigma^+))/\sigma(pp\to
K^+p\Lambda) = (4\pm2)\times 10^{-3}$, from which we deduce that the
total cross section for $\Sigma^+$ production at 1.826~GeV
($\varepsilon = 13$~MeV) is below 45~nb at the 98\% confidence level. 
This upper bound already excludes the COSY-11 result~\cite{ROZ06} by 
a very large margin. Even if the ratio of the $\Sigma^+/\Sigma^0$ 
production cross sections were a factor of six, as allowed by the 
triangle inequality at $\varepsilon=129$~MeV, the resulting 
distribution would greatly overestimate the data shown in Fig.~\ref{fig:Ratio}.

%
%
\subsection{Inclusive cross sections}
\label{Inclusive_sigma}

\begin{figure}[h!]
\begin{center}
\includegraphics[scale=0.7]{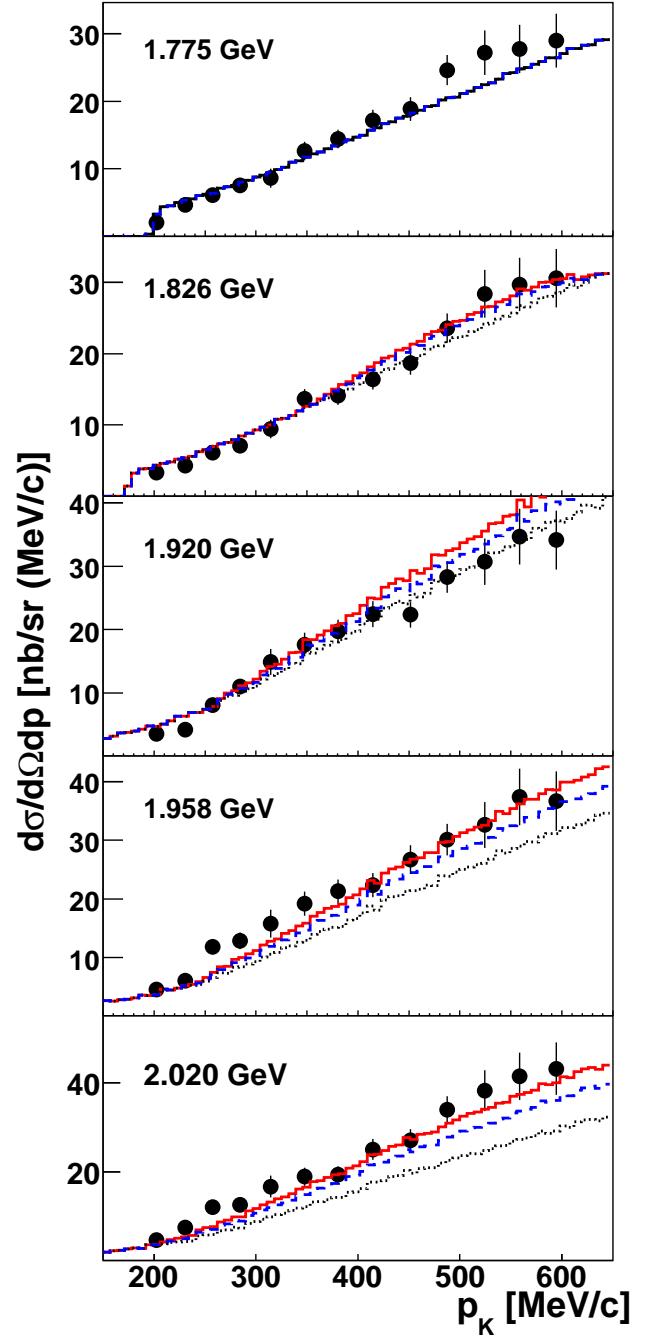}
\end{center}
\caption{(Color online)
Laboratory differential cross sections for inclusive $K^+$ production 
at the five measured energies, evaluated over the angular region 
$\vartheta_K < 4^{\circ}$. The overall uncertainty of 7\% coming from 
the luminosity has not been included. The experimental results are 
compared with Monte Carlo simulations; $\Lambda$ production 
(black dotted line), $\Lambda+\Sigma^0$ production (blue dashed line),
 $\Lambda+\Sigma^0+\Sigma^+$ production (red solid line). \label{fig:fDiffCr}}
\end{figure}

The average momentum $p_{K}$ of the $K^+$ meson, detected in each of
the telescopes, as well as the corresponding values of the inclusive
double-differential cross sections, evaluated using
Eq.~\eqref{eq:d2sigmadOmdP}, are presented in Table~\ref{tab:DDCr}.
These have then been summed and averaged over the kaon angular range
$\vartheta_{K^+}{<}4^{\circ}$ to give the inclusive cross sections
shown in Fig.~\ref{fig:fDiffCr} for the five beam energies studied.
The data are compared with simulations of $\Lambda$, $\Sigma^0$ and
$\Sigma^+$ production discussed in Sec.~\ref{data_and_models}, where
the individual contributions from the three reaction channels have
been normalized to the total cross sections deduced from the $K^+p$
and $K^+\pi^+$ correlation data, as described in
Sec.~\ref{Discussion}. Although there is fair agreement between the
simulations and the data for all five energies, it has to be stressed 
there is considerable ambiguity in the dominant $\Lambda$ production
cross section because in this case the reaction is measured far above
 threshold. The COSY-TOF results show that there are then strong
deviations from phase space for both the angle and momentum of the
$K^+$~\cite{SAM06}.

\begin{table*}[ht]
\begin{center}
\begin{tabular}{|c|c|c|c|c|c||c|}
\hline
$p_K$   &\multicolumn{5}{c|}{$d\sigma/ d\Omega dp$}&\\
MeV/$c$ &\multicolumn{5}{c|}{nb/(sr\,MeV/$c$)}&$R(1.826/1.775)$\\
\hline
231 & $\phantom{1}4.7\pm0.5$ & $\phantom{1}4.3\pm0.5$ & $\phantom{1}4.2\pm0.5$ & $\phantom{1}6.1\pm0.7$ & $\phantom{1}7.5\pm0.9$ & $0.91 \pm 0.02$\\
258 & $\phantom{1}6.1\pm0.7$ & $\phantom{1}6.1\pm0.7$ & $\phantom{1}8.1\pm0.9$ & $11.8\pm1.3$ & $12.0\pm1.3$ & $0.98 \pm 0.02$\\
285 & $\phantom{1}7.6\pm0.9$ & $\phantom{1}7.1\pm0.8$ & $\phantom{1}11.0\pm1.2$ & $12.8\pm1.4$ & $12.6\pm1.4$ & $0.94 \pm 0.02$\\
315 & $\phantom{1}8.6\pm1.4$ & $\phantom{1}9.4\pm1.3$ & $\phantom{1}14.9\pm2.1$ & $15.8\pm2.4$ & $16.7\pm2.5$ & $0.90 \pm 0.04$\\
348 & $12.6\pm1.4$ & $13.6\pm1.4$ & $17.6\pm1.9$ & $19.2\pm2.1$ & $18.9\pm2.0$ & $1.03 \pm 0.02$\\
381 & $14.4\pm1.4$ & $14.2\pm1.3$ & $19.7\pm1.8$ & $21.3\pm2.0$ & $19.4\pm1.9$ & $0.93 \pm 0.02$\\
415 & $17.1\pm1.6$ & $16.4\pm1.5$ & $22.4\pm2.0$ & $22.4\pm2.1$ & $25.0\pm2.3$ & $0.95 \pm 0.02$\\
452 & $18.9\pm1.7$ & $18.7\pm1.7$ & $22.3\pm2.0$ & $26.7\pm2.5$ & $27.1\pm2.5$ & $1.00 \pm 0.02$\\
488 & $24.6\pm2.2$ & $23.6\pm2.1$ & $28.3\pm2.5$ & $30.1\pm2.7$ & $34.0\pm3.0$ & $0.99 \pm 0.02$\\
525 & $27.2\pm3.3$ & $28.4\pm3.3$ & $30.7\pm3.7$ & $32.6\pm3.9$ & $38.3\pm4.6$ & $1.01 \pm 0.02$\\
559 & $27.8\pm3.6$ & $29.7\pm3.7$ & $34.7\pm4.4$ & $37.4\pm4.8$ & $41.4\pm5.3$ & $1.01 \pm 0.02$\\
595 & $29.0\pm4.0$ & $30.6\pm4.1$ & $34.2\pm4.6$ & $36.6\pm5.1$ & $43.2\pm5.9$ & $0.99 \pm 0.02$\\
\hline
$T_p$ [GeV] & 1.775 & 1.826 & 1.920 & 1.958 & 2.020 &\\
\hline
\end{tabular}
\end{center}
\caption{The $K^+$ laboratory double-differential cross sections measured 
at five beam energies are integrated over $\vartheta_{K^+}<4^{\circ}$ and 
momentum bins of width $\pm12$~MeV/$c$. The errors do not include the systematic
uncertainty of 7\% coming from the normalization. The final column shows 
the ratio of $K^+$ count rates measured at 1.826 and 1.775~GeV
in the interval $\vartheta_{K^+}<12^{\circ}$ and $\pm12$~MeV/$c$ momentum bins.
\label{tab:DDCr}
}
\end{table*}
%
%

\subsection{Kaon-proton coincidence measurements}
\label{Kp_results}

\begin{figure}[htb]
\begin{center}
\includegraphics[scale=0.6]{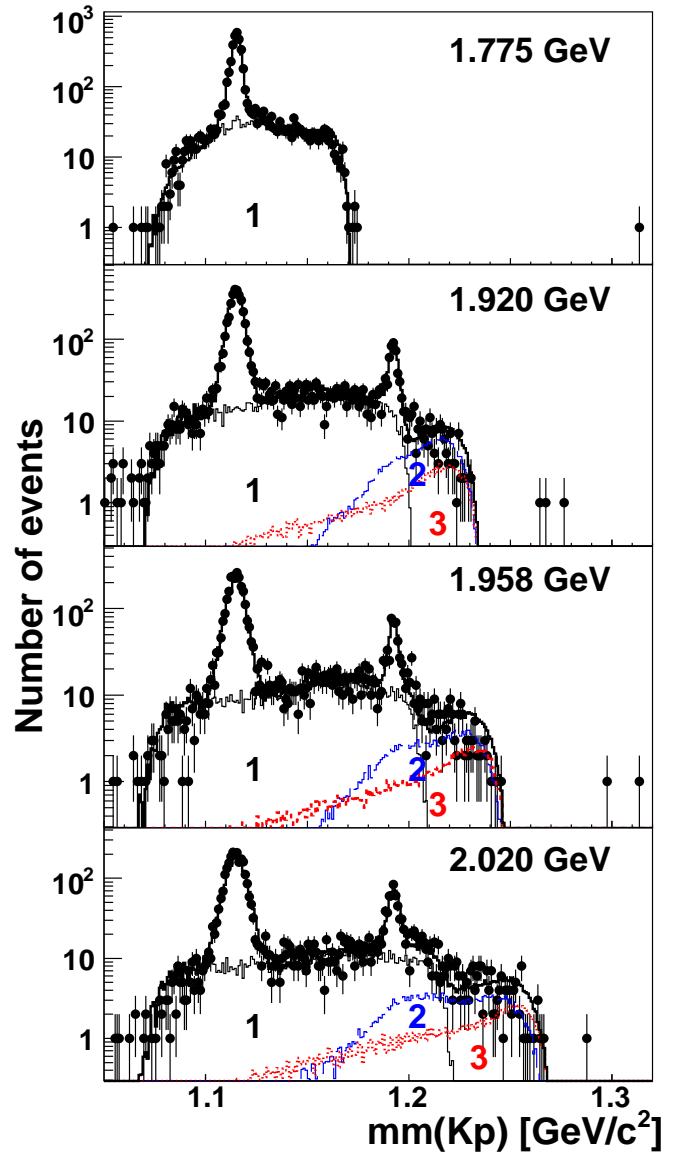}
\end{center}
\caption{(Color online) The $K^+p$ missing-mass spectra measured at 
four beam energies (closed symbols) compared to simulations normalized 
to the total cross sections listed in Tables~\ref{tab:ResTot} and 
\ref{tab:ResSigPlus}. In addition to the protons from $pp\to K^+p\Lambda/\Sigma^0$, 
these include contributions from indirect $K^+p$ correlations 
involving decay protons arising from the production of the $\Lambda$ 
(histogram 1), $\Sigma^0$ (histogram 2 (blue)), and $\Sigma^+$ 
(histogram 3 (red)). As explained in the
text, the $K^+p$ correlation data were unreliable at 1.826~GeV 
and so results are not shown for this energy. \label{fig:fKpmm}}
\end{figure}

The $K^+p$ missing-mass spectra measured at four beam energies are
presented in Fig.~\ref{fig:fKpmm}. Due to technical problems in the
preparation of the $K^+p$ correlation trigger, the spectrum measured
at $1.826$~GeV could not be reconstructed reliably and is therefore
not shown. However, this problem does not affect either the $K^+$
inclusive or the $K^+\pi^+$ correlation data collected at this
energy.

Figure~\ref{fig:fKpmm} shows clear peaks associated with the $pp\to
K^+p\Lambda$ and $pp\to K^+p\Sigma^0$ production processes, where the 
``$p$'' corresponds to the proton that is detected in ANKE. For events 
that lie within the $\Sigma^0$ peak, the distributions of $K^+$ 
missing masses are consistent with predictions based upon a phase-space 
production model, with no evidence for any strong $\Sigma^0p$ final 
state interaction.

In addition to the direct protons, there are large ``indirect'' physics 
backgrounds, where the $K^+$ is measured in coincidence with a proton 
that arises from the decay of one of the hyperons produced. The values 
of the relevant decay fractions are given in Table~\ref{tab:tRCh}. 
Both types of contribution from the three reaction channels have been 
simulated and traced through the GEANT4 model of the ANKE spectrometer. 
In view of their proximity to the thresholds, as well as the evidence 
discussed above, the contributions to the spectra from the two $\Sigma$ 
channels have been estimated using simple phase-space descriptions. 
However, $\Lambda$ hyperon production, which dominates the 
distributions, has been simulated using the different models 
mentioned in Sec.~\ref{data_and_models}. In all cases the protons 
that were produced in the decay of the three hyperons were assumed 
to be distributed according to phase space.

The numbers of detected $\Lambda$ and $\Sigma^0$ events could be fixed reliably
from fits to the experimental peaks and do not depend significantly on the model
used for the $\Lambda$ production. The indirect protons from $\Lambda$ decay lead
to a missing-mass distribution that dies off well before the kinematic limit, as can
be judged from the simulations shown in Fig.~\ref{fig:fKpmm}. Above the $\Lambda$
limit there are contributions coming from protons from $\Sigma^0$ and $\Sigma^+$
decay and these give rise to distributions with somewhat different shapes. The
numbers of $\Sigma^0$ events expected in this region can be estimated from the population
in the $\Sigma^0$ peak and, by subtraction, the numbers of $\Sigma^+$ events can
be estimated. The resulting values for the $\Sigma^+$ production cross section are
$0.25\pm0.05$, $0.41\pm0.09$, and $1.09\pm0.16$~$\mu$b at 1.920, 1.958 and 2.020~GeV,
respectively. Only statistical error bars are quoted and it is hard to obtain
robust estimates of the systematic uncertainties due to the proximity of the
kinematic limit. Conservative upper limits on the $\Sigma^+$ production cross section
can be derived if one assumes that the whole contribution to the spectra above
the $\Lambda$ limit comes from $\Sigma^+$ production. This extreme hypothesis
gives values of 0.51, 0.83, and 1.62~$\mu$b at the three energies. 
These results are given in Table~\ref{tab:ResSigPlus} along with those 
obtained from the $K^+\pi^+$ analysis.

The histograms in Fig.~\ref{fig:fKpmm}
present our estimations for all three reaction channels, normalized
to the total cross sections listed in Tables~\ref{tab:ResTot} and \ref{tab:ResSigPlus}. If the
value of the $\Sigma^+$ cross section at 1.958~GeV were as large as
that advocated by the COSY-11 collaboration~\cite{ROZ06}, this would
require boosting the red histogram in Fig.~\ref{fig:fKpmm} by about
two orders of magnitude.
%
%

\subsection{Kaon-pion coincidence measurements}
\label{KPi_results}

Below the threshold for the $pp \to K^+ n \Lambda \pi^+$ reaction
($T_{\text{thr}}=1.975$~GeV), the $K^+\pi^+$ coincidences arising
from the hyperon decay $\Sigma^+ \to n \pi^+$ are a unique signature
of the $pp \to K^+n\Sigma^+$ reaction. The background from the
$K^+n\Lambda \pi^+$ channel will remain negligible also at 2.02~GeV
because of the very small phase space and the fact that at the much
higher energy of $2.88$~GeV it comprises only 4\% of the $\Sigma^+$
production cross section~\cite{LOU61}. The smallness of the $K^+ n
\Lambda \pi^+$ production rate was also confirmed by our studies at
$2.16$~GeV, where no evidence for this channel was found in the
kaon-proton and kaon-pion coincidence spectra~\cite{VAL07}.

\begin{figure}[htb]
\begin{center}
\includegraphics[scale=0.55]{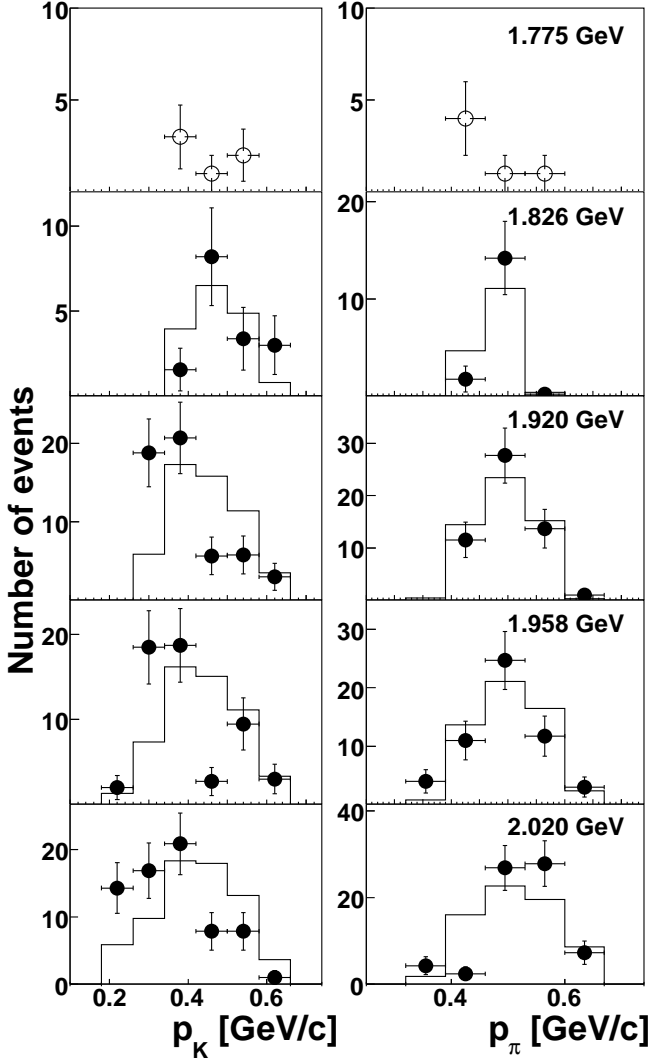}
\end{center}
\caption{Spectra of the kaon and pion momenta from identified
$K^+\pi^+$ pairs measured at four energies above the $\Sigma^+$ 
threshold and one below. The pions are taken over the full
 horizontal acceptance of ANKE ($\approx 15^{\circ}$) whereas the kaons are cut
at 12$^{\circ}$. Random background has been subtracted bin-by-bin
from the above-threshold data by scaling the 1.775~GeV results by
 the relative luminosity. The spectra are compared to phase-space
simulations of the $pp\to K^+n\Sigma^+$ reaction channel.
\label{fig:fKPiMomSp}}
\end{figure}

The measured kaon and pion momentum spectra from $K^+\pi^+$
coincidence events are presented in Fig.~\ref{fig:fKPiMomSp}, where
they are compared with phase-space simulations of the $pp\to K^+ n
\Sigma^+$ reaction. The number of random coincidences that survive
the analysis procedure discussed in Sec.~\ref{Experiment} can be
judged from the spectra measured at $T_p=1.775$~GeV, which is below
the $\Sigma^+$ threshold. The kinematically allowed region in the 
two dimensional plane of the $K^+$ and $\pi^+$ momenta has been
determined from simulations carried out at each energy. The resulting 
kinematical cuts have been applied, not only on the distributions
above threshold, but also on the 1.775~GeV data. To compensate for
the random background, the number of events estimated from the
subthreshold data has been subtracted from the data at the different
energies, taking into account the relative total luminosities.

\begin{table}[ht]
\begin{center}
\begin{tabular}{|c|c|c|c|c|c|}
\hline
$\varepsilon$&$L_{\text{tot}}$&$N_{\text{tot}}$&$\varepsilon_{K^{+}\pi^{+}}^{\text{acc}}$\\
MeV          & pb$^{-1}$      & event          & \%\\
\hline
13& 3.53&16&2.3\\
47& 2.17&54&0.6\\
60& 1.49&54&0.5\\
82& 1.40&69&0.1\\
\hline
\end{tabular}
\end{center}
\caption{Input information for the $\Sigma^+$ total cross section
estimation on the basis of the $K^+\pi^+$ correlation data. Values 
are given for the excess energy ($\varepsilon$),
total luminosity ($L_{\text{tot}}$), accumulated number of $K^+\pi^+$ 
coincidences after background subtraction ($N_{\text{tot}}$), and
total geometrical acceptance used
($\varepsilon_{K^{+}\pi^{+}}^{\text{acc}}$). \label{tab:tKPiInfo} }
\end{table}

After subtracting the background bin-by-bin and correcting for
efficiencies, the $K^+$ and $\pi^+$ momentum spectra measured above
the $\Sigma$ threshold are compared in Fig.~\ref{fig:fKPiMomSp} with
phase-space simulations. Although the numbers of events are small 
and the fluctuations large, both the $K^+$ and $\pi^+$ distributions are
in reasonable agreement with the simulations, within the uncertainties
 of the background subtraction procedure. The number of
$K^+\pi^+$ coincidences extracted from the spectra, together with
estimates of the overall geometrical acceptance (see Tab.~\ref{tab:tKPiInfo}), 
including all the efficiencies, lead to evaluations of the total 
cross sections on the basis of Eq.~\eqref{eq:totcr}. These are presented in
Table~\ref{tab:ResSigPlus}, where the very satisfactory agreement 
with the results derived from the $K^+p$ correlation data is demonstrated.
%
%
\section{Total cross sections for hyperon production}
\label{Discussion}%

The results obtained for the total cross sections of hyperon
production in proton-proton collisions are presented in
Table~\ref{tab:ResTot} and \ref{tab:ResSigPlus}. The values in the $\Lambda$ and $\Sigma^0$
cases are extracted from the fits to the two peaks in the $K^+p$ correlation spectra
of Fig.~\ref{fig:fKpmm}. Since the data were taken far from the
$\Lambda$ threshold where the ANKE acceptance is low, the
uncertainties here are dominated by the ambiguities in the models
used to describe the differential distributions. Any model dependence 
is far weaker for $\Sigma^0$ production and the contributions from other 
systematic effects, including luminosity, are of a similar size.

\begin{table}[ht]
\begin{center}
\begin{tabular}{|c|c|c|c|c|c|}
\hline
Hyperon    & $T_{p}$ & $\varepsilon$  & $\sigma$ & $\Delta_{\text{stat}}$ & $\Delta_{\text{syst}}$\\
           & GeV     &MeV             &$\mu$b  & \% &\%\\
\hline
           & 1.775 & $\phantom{1}70\pm1$  & 6.1  &2 & 18\\
$\Lambda$  & 1.920 & $122\pm1$ & 16.9 &2 & 21\\
           & 1.958 & $135\pm1$ & 16.1 &2 & 21\\
           & 2.020 & $157\pm1$ & 18.0 &2 & 31\\
\hline
           & 1.920 & $\phantom{1}45\pm1$  & 0.24 &5 & 14\\
$\Sigma^0$ & 1.958 & $\phantom{1}58\pm1$  & 0.51 &5 & 14\\
           & 2.020 & $\phantom{1}80\pm1$  & 1.30 &5 & 14\\
\hline
\end{tabular}
\end{center}
\caption{Total cross sections for $\Lambda$ and $\Sigma^0$ production 
measured in proton-proton collisions. \label{tab:ResTot}}
\end{table}

\begin{table}[ht]
\begin{center}
\begin{tabular}{|c|c||c|c|c||c|c|c|}
\hline
$T_{p}$ & $\varepsilon$  & $\sigma_{K\pi}$ & $\Delta_{\text{stat}}$ & $\Delta_{\text{syst}}$& $\sigma_{Kp}$ & $\Delta_{\text{stat}}$ & Upper limit\\
GeV     &MeV             &$\mu$b  & \% &\% &$\mu$b & \% & $\mu$b\\
\hline
 1.826 & $13\pm1$ & 0.011 & 38 & 17 & --& --& 0.045\\
 1.920 & $47\pm1$ & 0.20  & 16 & 17 & 0.25 & 21 & 0.51\\
 1.958 & $60\pm1$ & 0.39  & 15 & 17 & 0.41 & 23 & 0.83\\
 2.020 & $82\pm1$ & 0.76  & 14 & 17 & 1.09 & 15 & 1.62\\
\hline
\end{tabular}
\end{center}
\caption{Total cross sections for $pp\to K^+n\Sigma^+$ production
 measured in this experiment. The values in the column marked 
$\sigma_{K\pi}$ were obtained from the study of the $K^+\pi^+$ 
coincidence data whereas those in column $\sigma_{Kp}$ followed 
from the study of the high $K^+p$ missing-mass spectra of 
Fig.~\ref{fig:fKpmm}. As discussed in the text, it is very hard 
to quantify satisfactorily the systematic uncertainties in this case. 
The upper limit quoted in the last column at 1.826~GeV was derived 
from the study of inclusive $K^+$ production above and below threshold 
shown in Fig.~\ref{fig:Ratio}. At the higher energies, the upper 
limits were obtained by assuming that all the events in 
Fig.~\ref{fig:fKpmm} that fell above the kinematic limit for 
the $\Lambda$ were associated with $\Sigma^+$ production. \label{tab:ResSigPlus}}
\end{table}

Table~\ref{tab:ResSigPlus} shows the values of the $\Sigma^+$ 
production cross section extracted from both the $K^+\pi^+$ and 
$K^+p$ correlation data shown in Figs.~\ref{fig:fKPiMomSp} and \ref{fig:fKpmm}, respectively.
As previously remarked, experimental difficulties at 
$T_p=1.826$~GeV made the $K^+p$ correlation data unreliable at 
this energy, though this had no influence on the $K^+\pi^+$ 
measurement. The principal errors in the $K^+\pi^+$ case arise 
mainly from the low statistics. These are much better in the 
$K^+p$ coincidence measurement but the difficulties associated 
with evaluating the acceptance at the kinematic limit makes it 
hard to quantify the systematic errors. Despite these uncertainties, 
the agreement between the two sets of measurements is very reassuring. 
Upper limits on the $\Sigma^+$ production cross section are also 
quoted in Table~\ref{tab:ResSigPlus}. That at the lowest energy 
was extracted at the 98\% confidence level from the comparison 
of the inclusive $K^+$ production data at 1.826 and 1.775~GeV 
shown in Fig.~\ref{fig:Ratio}. At the three higher energies 
the upper bounds were determined from the assumption that 
all the events above the $\Lambda$ limit in the $K^+p$ 
missing-mass spectra of Fig.~\ref{fig:fKpmm} corresponded 
to $\Sigma^+$ production. At 1.958~GeV this is about a factor 
of 50 below the result of the COSY-11 collaboration~\cite{ROZ06}. 
It can be seen from the tables that the $\Sigma^+$ production 
cross sections are only a little less than those of $\Sigma^0$ 
but that the difference is within the error bars.

\begin{figure}[!h]
\begin{center}
\includegraphics[scale=0.5]{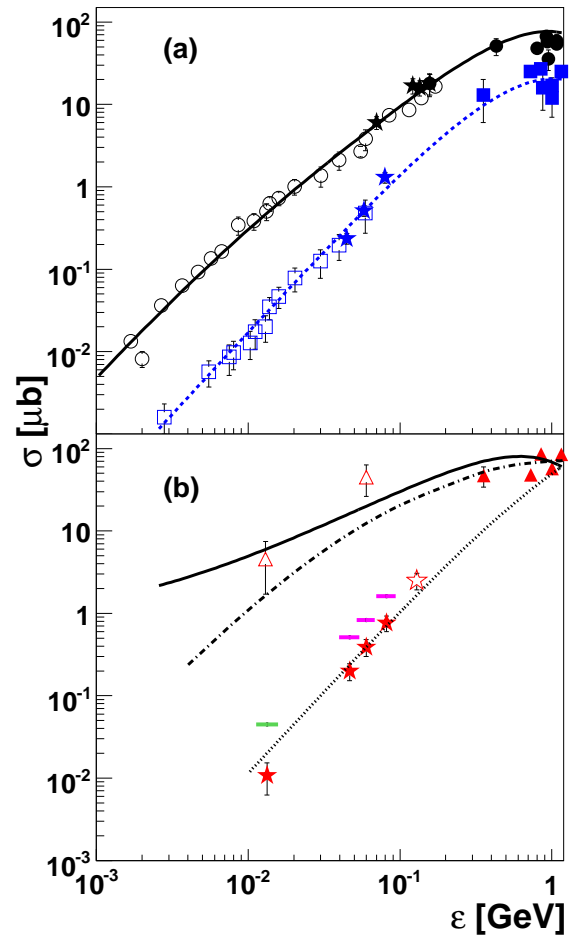}
\end{center}
\caption{(Color online) Total cross sections for hyperon production
in proton-proton collisions as functions of excess
 energy obtained in this experiment and compared to published data. Panel (a): Results
for $\Lambda$ (black stars) and $\Sigma^0$ (blue stars). The data
points obtained in other experiments at COSY and other 
facilities are shown by open and closed symbols (circles for $\Lambda$ and squares
for
$\Sigma^0$)~\cite{BAL96,BIL98,SEW99,KOW04,SAM06,VAL07,LOU61,ABD04,ABD07,BAL88}.
The lines represent parameterizations of the world data from
Ref.~\cite{SIB06}. Panel (b): $\Sigma^+$ production. Closed stars
represent the current data set obtained from the $K^+\pi^+$ coincidence spectra.
The open star corresponds to our previous measurement~\cite{VAL07}. 
The conservative upper limits given in Table~\ref{tab:ResSigPlus} 
are shown by the four short horizontal lines. The COSY-11 
results~\cite{ROZ06} are shown by open triangles
while the closed ones are taken from a compilation of higher energy
data~\cite{BAL88}. The dotted line corresponds to a pure phase-space
variation with a normalization chosen to pass through our central
energies. The solid line is a parameterization that includes a 
strong $n\Sigma^+$ FSI and energy dependent matrix element~\cite{CAO08}. 
The broken line is taken from the microscopic calculation of
Ref.~\cite{XIE07}. \label{fig:fTotCrEps}}
\end{figure}

The results for all three hyperons are presented in graphical form 
in Fig.~\ref{fig:fTotCrEps}. It is seen from this that our values 
of the production cross sections for both $\Lambda$ and $\Sigma^0$ 
are consistent with the world data. However, it has to be stressed 
once again that a large systematic error bar had to be ascribed 
to the $\Lambda$ value because of the small ANKE acceptance. 
As is well known, it is important to introduce the $\Lambda p$ 
final state interaction to describe the energy dependence of 
the $pp\to K^+p\Lambda$ total cross section. On the other hand 
it is seen that the $pp\to K^+p\Sigma^0$ variation is much 
closer to phase space, though our data might give a possible 
hint of some small repulsive FSI.

The $\Sigma^+$ production cross section in $pp$ collisions shows no
sign of any threshold enhancement of a type that could explain the
large values found by the COSY-11 collaboration~\cite{ROZ06} and,
indeed, their values are already completely excluded by the upper 
bounds  provided by the $K^+$ inclusive and $K^+p$ correlation 
data of Figs.~\ref{fig:Ratio},
\ref{fig:fDiffCr} and \ref{fig:fKpmm}. The theoretical
models~\cite{CAO08,XIE07} whose predictions are shown in
Fig.~\ref{fig:fTotCrEps}b were motivated by the COSY-11 data 
and fail completely to reproduce our results.

The ratio of $\Sigma^+$ to $\Sigma^0$ production seems to 
lie between about a half and one, though the difference is 
possibly within the error bars. The energy dependence shown in
Fig.~\ref{fig:fTotCrEps}b seems a little steeper than that suggested on
phase-space grounds and, if this is so, it could be an 
indication for a damping effect near threshold due to a $\Sigma^+n$ final state
interaction. There is ample evidence from the study of the threshold
cusp in $K^-d\to \pi^-\Lambda p$~\cite{TAN69} that the $\Lambda p$
and $\Sigma N$ channels are strongly coupled. The enhancement in the
$K^+$ missing mass in the inclusive $pp\to K^+X$ data at the $\Sigma
N$ threshold~\cite{SIE94,HIRES} is a possible sign that flux 
is being removed from the $pp\to K^+N\Sigma$ channels in this region.

%
%
\section{Conclusions}
\label{Conclusions}

We have reported measurements of the $pp\to K^+ n\Sigma^+$ reaction
at four energies close to threshold using two techniques. Instead of identifying the
neutron, as was attempted at COSY-11~\cite{ROZ06}, the $K^+$ was
detected in coincidence with the $\pi^+$ that came from the decay of
the $\Sigma^+$ hyperon. Below the threshold for $K^+n\Lambda\pi^+$
production, this decay is the only source of such correlations. The
numbers of random $K^+\pi^+$ coincidences were small and could be
estimated from data taken below the threshold for $\Sigma^+$
production. These results were confirmed within error bars from the
study of the high end of the $K^+p$ missing-mass spectra, though it
has to be stressed that systematic effects are harder to control in 
this measurement. Nevertheless, conservative upper limits on $\Sigma^+$ 
production could be obtained by assuming that the $\Sigma^0$ events did
not populate the high $K^+p$ missing-mass distributions. 
In complete contrast to the COSY-11 results~\cite{ROZ06},
we found cross sections for the $pp\to K^+ n\Sigma^+$ reaction to be
even a little smaller than those of $pp\to K^+ p\Sigma^0$.

The large COSY-11 $\Sigma^+$ production rates could also be excluded
by the study of inclusive $K^+$ production~\cite{SIB07}. However,
this does not provide the best estimates of the 
$pp \to K^+n\Sigma^+$ cross section because the $K^+$ spectrum is dominated by the
$K^+p\Lambda$ channel, unless $\Sigma^+$ production is anomalously
large near threshold~\cite{ROZ06}. The uncertainty in the $\Sigma^+$
cross section deduced from such data cannot therefore be 
smaller than the sum of the uncertainties in the models describing $\Lambda$ and
$\Sigma^0$ production, which is why these data only provide upper
limits on $\Sigma^+$ production.

The data on $\Lambda$ production were taken far above threshold 
where the acceptance of ANKE for the $pp\to K^+p\Lambda$ 
reaction is small. The predicted numbers of inclusive $K^+$ depend sensitively upon the
variation of the production with angle and momentum. In addition to
the strong distortion due the $\Lambda p$ final state 
attraction, the COSY-TOF data~\cite{SAM06} show significant angular variation. As a
consequence, one cannot extract total cross sections for $\Lambda$
production in a model-independent way from an apparatus with a small
acceptance in angle. The situation would be even more uncertain if
the strong $\Lambda p\rightleftharpoons \Sigma N$ channel coupling~\cite{TAN69} gave
rise to a rapid $K^+$ momentum dependence. Our aim was merely
to show that, with reasonable model assumptions, we could achieve
total cross sections that were not inconsistent with world data.

The experimental situation regarding acceptance is far more
comfortable for the $K^+p\Sigma^0$ channel in this respect because the results were
obtained not far above threshold. Furthermore, it is believed that
the $\Sigma^0p$ final state interaction is much weaker than that of
the $\Lambda p$ and our data agree with this. The resulting 
values of the $pp\to K^+p\Sigma^0$ cross section shown in 
Fig.~\ref{fig:fTotCrEps}a are completely consistent within 
error bars with published data and this supports the validity 
of our experiment and its analysis.

Although the $K^+\pi^+$ coincidence method leads to only small
numbers of events within our acceptance, these are very clean and so
the total cross sections presented in Fig.~\ref{fig:fTotCrEps}b have
fewer systematic uncertainties than those coming from the $K^+p$ correlation data. 
It may therefore be interesting to note that, taken together with our 2.16~GeV
point~\cite{VAL07}, the data seem to indicate a somewhat steeper rise than
expected on the basis of phase space. This is definitely the case if
the data have to join smoothly on to the bubble chamber measurements
shown in Fig.~\ref{fig:fTotCrEps}.

Attractive final state interactions, such as those in the $pp$ or
$\Lambda p$ systems, give rise to a very strong energy dependence
because they can generate virtual state poles in the production
amplitudes. This is not the case when there is repulsion between two
final particles and so this would, in general, lead to much smaller
effects. There is, however, a third type of final state interaction,
namely an absorptive one where the final wave function is damped at
small distances rather than being repelled. In the present case this
could arise from the coupling of the $\Sigma N$ and the $\Lambda p$
channels, which is known to be very strong at low
energies~\cite{TAN69}. It is therefore possible that flux from the
$K^+N\Sigma$ channels is lost to feed the $K^+p\Lambda$ channel.

If the coupled channel final state interaction is significant, it
would only be relevant in the isospin $I=\frac{1}{2}$ state and this
would have twice as big an effect on $K^+n\Sigma^+$ production
compared to $K^+p\Sigma^0$. It is perhaps noteworthy that the
parameterization of the world data for $pp \to K^+p\Sigma^0$ 
shown in Fig.~\ref{fig:fTotCrEps}a do not seem to indicate as steep a rise as
for $pp \to K^+n\Sigma^+$, though this is not the case for our own
$\Sigma^0$ data.

As stressed in Sec.~\ref{data_and_models}, there are isospin links
between the amplitudes for $pp\to K^+p\Sigma^0$, $pp\to
K^+n\Sigma^+$, and $pp\to K^0p\Sigma^+$, but these all refer to an
initial $NN$ isospin-one state. The isospin-zero amplitude 
for $pn\to K^+p\Sigma^-$ is independent of these and so extra help for the
modeling might be provided by measurements in this channel. This
will be possible at COSY-ANKE through the study of quasi-free
production on the deuteron with the detection of the spectator
proton~\cite{VPK}.

The existing theoretical models for $pp\to K^+n\Sigma^+$
production~\cite{CAO08,XIE07} bear no relation to our results. 
If the reaction were driven by the exchange of non-strange 
mesons exciting a $\Delta$ isobar~\cite{XIE07}, this would strongly favor the
production of the $\Sigma^+$ compared to the $\Sigma^0$, which is in
conflict with the data. It would also not contribute to the
isospin-zero production of $\Sigma^-$.  Further theoretical work is
clearly needed to understand the full data set.\\
%
%
\begin{acknowledgments}
We wish to thank the COSY team, especially D.~Prasuhn and B.~Lorentz
for their support throughout this experiment. Useful 
discussions with A.~Sibirtsev and other members of the ANKE Collaboration are
gratefully acknowledged. This work has been partially supported by
BMBF, Russian Academy of Science, the JCHP FFE, and the HGF-VIQCD.
\end{acknowledgments}
%
%
%


\begin{thebibliography}{99}
%
\bibitem{BAL96} J.T.~Balewski \emph{et al.}, Phys.\ Lett.\ B \textbf{388}, 859 (1996);
Phys.\ Lett.\ B \textbf{420}, 211 (1998); Eur.\ Phys.\ J.\ A
\textbf{2}, 99 (1998).
%
\bibitem{BIL98} R.~Bilger \emph{et al.}, Phys. Lett. B \textbf{420}, 217 (1998).
%
\bibitem{SEW99} S.~Sewerin \emph{et al.}, Phys.\ Rev.\ Lett.\ \textbf{83}, 682 (1999).
%
\bibitem{EYR03} W.K.~Eyrich, Prog.\ Part.\ Nucl.\ Phys.\ \textbf{50}, 547 (2003).
%
\bibitem{KOW04} P.~Kowina \emph{et al.}, Eur.\ Phys.\ J.\ A \textbf{22}, 293 (2004).
%
\bibitem{SAM06} S.~Abd El-Samad \emph{et al.}, Phys.\ Lett.\ B \textbf{632}, 27
(2006).
%
\bibitem{VAL07} Yu.~Valdau \emph{et al.}, Phys.\ Lett.\ B \textbf{652}, 245 (2007).
%
\bibitem{ROZ06} T.~Ro\.zek \emph{et al.}, Phys.\ Lett.\ B \textbf{643}, 251 (2006).
%
\bibitem{LOU61} R.I.~Loutitt \emph{et al.}, Phys.\ Rev.\
\textbf{123}, 1465 (1961).
%
\bibitem{VAL09} Yu.~Valdau, Ph.D.\ thesis, University of Cologne (2009), available from
\url{www.fz-juelich.de/ikp/anke/en/theses.shtml}.
%
\bibitem{SIE94} R.~Siebert \emph{et al.}, Nucl.\ Phys.\ A
\textbf{567}, 819 (1994).
%
\bibitem{SIB07} A.~Sibirtsev, J.~Haidenbauer, H.-W.~Hammer, and U.-G.~Mei{\ss}ner,
Eur.\ Phys.\ J.\ A \textbf{32}, 229 (2007).
%
\bibitem{HIRES} A.~Budzanowski \emph{et al.}, \textit{High resolution
study of the $\Lambda p$ final state interaction in the reaction
$p+p\to K^++(\Lambda p)$}, submitted to Phys.\ Lett.\ B (2009).
%
\bibitem{SIB06} A.~Sibirtsev, J.~Haidenbauer, H.-W.~Hammer, and
S.~Krewald, Eur.\ Phys.\ J.\ A \textbf{27}, 269 (2006).
%
\bibitem{MET98} A.~Metzger, Ph.D.\ thesis, University of Erlangen--N\"urnberg
  (1998);
This and Refs.~\cite{HES00,FRI02,SCH03,KAR05,SCH03a} are 
available from \url{www.fz-juelich.de/ikp/COSY-TOF/publikationen}.
%
\bibitem{HES00} D.~Hesselbarth, Ph.D.\ thesis, University of Bonn (2000).
%
\bibitem{FRI02} M.~Fritsch, Ph.D.\ thesis, University of Erlangen--N\"urnberg (2002).
%
\bibitem{SCH03} W.~Schroeder, Ph.D.\ thesis, University of Erlangen--N\"urnberg (2003).
%
\bibitem{ABD04} M.~Abdel-Bary \emph{et al.}, Phys.\ Lett.\ B
\textbf{595}, 127 (2004).
%
\bibitem{ABD07} M.~Abdel-Bary \emph{et al.}, Phys.\ Lett.\ B
\textbf{649}, 252 (2007).
%
\bibitem{KAR05} L.~Karsch, Ph.D.\ thesis, University of Dresden
(2005).
%
\bibitem{SCH03a} P.~Sch\"{o}nmeier, Ph.D.\ thesis, University of
Dresden (2003).
%
\bibitem{BAR01} S.~Barsov \emph{et al.}, Nucl.\ Instrum.\
Methods Phys.\ Res.\ Sect.\ A \textbf{462}, 354 (2001).
%
\bibitem{MAI97} R.~Maier \emph{et al.}, Nucl.\ Instrum.\ Methods 
Phys.\ Res., Sect.\ A \textbf{390}, 1 (1997).
%
\bibitem{KHO99} A.~Khoukaz \emph{et al.}, Eur.\  Phys.\ J.\ D\ \textbf{5}, 275 (1999).
%
\bibitem{BUS02} M.~B\"uscher \emph{et al.}, Nucl.\ Instrum.\
Methods Phys.\ Res., Sect.\ A  \textbf{481}, 378 (2002).
%
\bibitem{DYM04} S.~Dymov \emph{et al.}, Part.\ Nucl.\ Lett.\ \textbf{2}, 40 (2004).
%
\bibitem{DYM00} S.N.~Dymov \emph{et al.}, Nucl.\ Instrum.\ Methods
Phys.\ Res., Sect.\ A \textbf{440}, 431 (2000).
%
\bibitem{AGO03} S.~Agostinelli \emph{et al.}, Nucl.\ Instrum.\ Methods Phys.\ 
Res., Sect.\ A \textbf{506}, 250 (2003);
\url{http://geant4.web.cern.ch/geant4}.
%
\bibitem{AMS08} C.~Amsler \emph{et al.}, Phys.\ Lett.\ B \textbf{687}, 1 (2008).
%
\bibitem{BAL88} A.~Baldini, V.~Flamino, W.G.~Moorhead, and D.R.O.~Morison,
Landolt-B\"{o}rnstein, New Series, Ed.\ H.~Schopper (Springer-Verlag, Berlin, 1988).
%
\bibitem{CAO08} Cao Xu, Lee Xi-Guo, and Wang Qing-Wu, Chin.\ Phys.\ Lett.\ B \textbf{25}, 888 (2008).
%
\bibitem{XIE07} Ju-Jun Xie and Bing-Song Zou, Phys.\ Lett.\ B \textbf{649}, 405 (2007).
%
\bibitem{TAN69} T.H.~Tan, Phys.\ Rev.\ Lett.\ \textbf{23}, 395 (1969).
%
\bibitem{ARN00} R.A.~Arndt \emph{et al.}, Phys.\ Rev.\ C\ \textbf{62} 034005 (2000); 
solution SP05 \url{http://gwdac.phys.gwu.edu}.
%
\bibitem{VPK} E.~Shikov, Diploma thesis,  St.~Petersburg State Polytechnical 
University (2009), available from
\url{www.fz-juelich.de/ikp/anke/en/theses.shtml}.
%
\end{thebibliography}
\end{document}